\documentclass[pra,reprint,showpacs,amsmath,amssymb,superscriptaddress]{revtex4-1}
\usepackage{graphicx}
\usepackage[dvipsnames]{xcolor}
\usepackage[colorlinks,bookmarks=false,citecolor=blue,linkcolor=red,urlcolor=blue]{hyperref}
\usepackage{multirow}

\begin{document}
\title{Nonlinear two-photon Rabi-Hubbard model: superradiance and
  photon/photon-pair Bose-Einstein condensate}
\author{Shifeng Cui}
\affiliation{Department of Physics, Beijing Normal University, Beijing
  100875, China}
\affiliation{Beijing Computational Science Research Center, Beijing
  100193, China}
\author{B. Gr\'emaud}
\affiliation{Aix Marseille Univ, Universit\'e de Toulon, CNRS, CPT, Marseille, France} 
\affiliation{MajuLab, CNRS-UCA-SU-NUS-NTU International Joint Research
  Unit, 117542 Singapore}
\affiliation{Centre for Quantum Technologies, National University of
  Singapore, 2 Science Drive 3, 117542 Singapore}
\author{Wenan Guo}
  \email{waguo@bnu.edu.cn}
\affiliation{Department of Physics, Beijing Normal University, Beijing
  100875, China}
\affiliation{Beijing Computational Science Research Center, Beijing 100193, China}
\author{G. G. Batrouni}
\email{george.batrouni@inphyni.cnrs.fr}
\affiliation{Universit\'e C\^ote d'Azur, CNRS, INPHYNI, France}
\affiliation{MajuLab, CNRS-UCA-SU-NUS-NTU International Joint Research
  Unit, 117542 Singapore}
\affiliation{Centre for Quantum Technologies, National University of
  Singapore, 2 Science Drive 3, 117542 Singapore} 
\affiliation{Department of Physics, National University of Singapore, 2
  Science Drive 3, 117542 Singapore}
\affiliation{Beijing Computational Science Research Center, Beijing 100193, China}

\begin{abstract}
We study the ground state phase diagram of a nonlinear two-photon
Rabi-Hubbard (RH) model in one dimension using quantum Monte Carlo
(QMC) simulations and density matrix renormalization group (DMRG)
calculations. Our model includes a nonlinear photon-photon interaction
term. Absent this term, the RH model has only one phase, the normal
disordered phase, and suffers from spectral collapse at larger values
of the photon-qubit interaction or inter-cavity photon hopping. The
photon-photon interaction, no matter how small, stabilizes the system
which now exhibits {\it two} quantum phase transitions: Normal phase
to {\it photon pair} superfluid (PSF) transition and PSF to single
particle superfluid (SPSF). The discrete $Z_4$ symmetry of the
Hamiltonian spontaneously breaks in two stages: First it breaks
partially as the system enters the PSF and then completely breaks when
the system finally enters the SPSF phase. We show detailed numerical
results supporting this, and map out the ground state phase diagram.
\end{abstract}


\maketitle

\section{Introduction}
Light-matter interaction is ubiquitous in nature and is, therefore,
the focus of much theoretical and experimental work. The relatively
simple Rabi model\cite{rabi36,rabi37} gives a good description of the
interaction of photons with a two level system (qubit) and recent
advances in manipulating the interactions of photons with single atoms
have resulted in its experimental realization using two-level atoms in
cavities\cite{haroche06} (cavity quantum electrodynamics, QED) or
Josephson junctions on solid state
chips\cite{wallraff04,chen07,lambert09,nataf10,viehmann11} (circuit
QED). In the strong coupling limit, $g/\omega \lesssim 0.1$, one can
exploit the random wave approximation (RWA) to justify ignoring the
counter-rotating (CR) terms resulting in a Hamiltonian which conserves
the exciton number (number of photons plus excited qubits) due to its
$U(1)$ symmetry. This results in the Jaynes-Cummings (JC)
model\cite{jaynes63} which can be solved exactly. Such cavities can be
arranged in a chain where the coupling between near neighbors can be
controlled via a tunable tunneling of the photon mode thus resulting
in the Jaynes-Cummings-Hubbard model\cite{schiro13} which describes
itinerant bosons (the photons) interacting with localized qubits. This
model has been shown to behave like the one dimensional Bose-Hubbard
model (BHM)\cite{batrouni90} with a superfluid phase (of excitons) and
incompressible Mott insulator (MI)
lobes\cite{greentree06,hartmann06,rossini07,koch08,hohenadler08,hartmann08,schmidt09,zhao08}.
The MI is approximately a product of single site states obtained from
a superposition of photons and excited atoms\cite{schmidt09} and
exhibits behavior similar to photon blockade\cite{birnbaum05} where
there is a finite energy gap opposing the addition of a photon.

In the experimentally
attainable\cite{niemczyk10,forndiaz10,chen17,forndiaz17,yoshihara17,kockum17},
ultra-strong coupling regime, $g/\omega\sim 1$, the CR terms can no
longer be ignored. Their restoration to the Hamiltonian reduces the
$U(1)$ symmetry to $Z_2$. Consequently, the number of excitons is no
longer conserved thus removing the possibility of a MI phase. The
phase diagram now consists of a disordered phase and an ordered
(coherent) one separated by a quantum phase transition in the
universality class of the two-dimensional Ising
model\cite{schiro13,zheng11,kumar13,flottat16}. The ordered phase,
therefore, exhibits a Bose-Einstein condensate (BEC) of photon. This
transition resembles the incoherent/coherent (normal/superradiant)
phase transition in the Dicke model\cite{hepp73,rotondo15}.

The two-photon Rabi model has also attracted much attention as new
systems are realized where multi-photon processes come into play. For
example, it has been used to describe second order processes in
Rydberg atoms in cavities\cite{bertet02} and quantum
dots\cite{stufler06,delvalle10}, and mechanisms have been proposed to
realize it in circuit QED\cite{felicetti18}. The two-photon model
undergoes spectral collapse where the Hamiltonian is no longer bounded
when the coupling exceeds a certain
value\cite{emary02,dolya09,travenec12,maciejewski15,travenec15,peng17,chen12,felicetti15}. In
the strong coupling regime, the CR terms can be ignored, as in the
one-photon case, and result in the two-photon JC model with $U(1)$
symmetry and conserved number of excitons. In the ultra-strong regime,
the CR terms are restored and the symmetry is reduced to $Z_4$. The
ground state of the many-body two-photon model was studied in the
context of the Dicke model\cite{dicke54} using mean
field\cite{garbe17} and QMC\cite{cui19}. Excellent agreement was found
between the two methods which showed the system to exhibit a quantum
phase transition between a normal (disordered) and superradiant
phase. Here, however, the term superradiant is used to indicate a
macroscopic change in the average number of photons but which remains
relatively small. This is in contrast with the one-photon case where
there is a very large number of photons in the superradiant phase
which form a BEC\cite{schiro13,zheng11,kumar13,flottat16}. QMC
simulations were also used to examine the two-photon JCH and Rabi
Hubbard (RH) models\cite{cui19}. It was found that the JCH model has
only one MI lobe with two excitons/cavity, unlike the one-photon case
where there is a succession of MI
lobes\cite{greentree06,hartmann06,rossini07,koch08,hohenadler08,hartmann08,schmidt09,zhao08}.
Furthermore, it was found that there are two superfluid (SF) phases; a
single photon SF phase (SPSF) and a photon pair SF phase (PSF). In the
former, the single particle Green function decays as a power
indicating the quasi-long range order for the photons. In the latter
case, however, the single particle Green function decays exponentially
while the photon pair Green function decays as a power indicating
quasi long range order for bound photon pairs but not for simgle
photons. When the CR terms are restored, the symmetry of the
Hamiltonian is reduced from $U(1)$ to $Z_4$ and it was found that the
two-photon RH model does not exhibit any quantum phase transitions; it
has only the disordered phase and the spectral collapse
region\cite{cui19}.

In this paper, we examine the possibility that non-linear photon terms
in the Hamiltonian could stabilize the system and allow the appearance
of quantum phase transitions in the RH model. Non-linear terms, {\it
  i.e.} photon-photon interactions, have attracted experimental and
theoretical interest as a means to generate topological photon
pairs\cite{gorlach17,mittal18,olekhno19} which have robust transport
properties. To this end we exploit the density matrix renormalization
group\cite{white92,white93} (DMRG) with open boundary conditions (OBC)
using the ALPS library\cite{alps}, and the stochastic Green
function\cite{rousseau08,rousseau08b} (SGF) quantum Monte Carlo
method, with periodic boundary conditions (PBC), to study the phase
diagram of the two-photon RH model with a nonlinear photon term. We
show that the photon-photon interaction term, stabilizes the system,
eliminating spectral collapse, and leads to the appearance of two
quantum phase transitions, the first is the transition from the
disordered phase to PSF phase where the $Z_4$ symmetry patially
breaks, the second is the transition between the PSF and the SPSF
phases where the $Z_4$ symmetry is completely broken. We show that in
the PSF phase, we have a BEC of photon pairs whereas in the SPSF phase
we have a BEC of photons.

The paper is organized as follows. In section {\bf II} we present the
model and discuss briefly the methods used to perform the numerical
calculations and simulations. In section {\bf III} we present and
sicuss our results, and in section {\bf IV} we discuss our
conclusions.

\section{Model and method}

The one-dimensional nonlinear two-photon Rabi-Hubbard (RH) model we
study is governed by the Hamiltonian,
\begin{eqnarray}
\nonumber
H_{RH}&=& -J \sum_{i=1}^N \left ( {\hat a}^\dagger_i {\hat 
a}^{\phantom\dagger}_{i+1} + h.c. \right ) + 
\sum_{i=1}^N
\left (\omega {\hat a}^\dagger_i {\hat a}^{\phantom\dagger}_i + 
\omega_q \sigma^+_i \sigma^-_i  \right )\\
&&+ g\sum_{i=1}^N \left ( \sigma^+_i  + \sigma^-_i \right ) \left
({\hat a}^2_i+{\hat a}^{\dagger 2}_i \right )+U\sum_i ({\hat
  a}^\dagger_i {\hat a}^{\phantom\dagger}_i)^2,
\label{rabihub}
\end{eqnarray}
where $N$ is the number of sites (or cavities) and ${\hat
  a}^{\phantom\dagger}_i$ (${\hat a}^\dagger_i$) is the photon
destruction (creation) operator in the $i$th cavity. Photon tunneling
between sites is governed by the hopping parameter $J$. The photon
frequency is $\omega$ and the qubit energy spacing is $\omega_q$;
$\sigma^z_i$ and $\sigma^x_i = \sigma^+_i + \sigma^-_i$ are the Pauli
matrices acting on the $i$th qubit, $\sigma^+_i$ ($\sigma^-_i$) is the
corresponding raising (lowering) operator. The last term in
Eq.(\ref{rabihub}) describes the onsite photon-photon
interaction. Ignoring the CR terms in Eq.~(\ref{rabihub}),
$(\sigma^+_i{\hat a}^{\dagger 2}_i + \sigma^-_i {\hat a}^2_i$), leads
to the Jaynes-Cummings-Hubbard (JCH) model where the system is
invariant under the generalized rotation operator,
\begin{equation}
 {\cal R}(\theta) = {\rm exp} \left (i \theta {\hat a}^\dagger_j 
{\hat a}^{\phantom\dagger}_j 
+ i2 \theta \sigma^+_j \sigma^-_j \right ),
\label{rotop}
\end{equation}
with ${\cal R}(\theta)^\dagger {\hat a}_j {\cal
  R}(\theta)^{\phantom\dagger}={\rm e}^{i\theta} {\hat a}_j$, and
${\cal R}(\theta)^\dagger \sigma^-_j {\cal
  R}(\theta)^{\phantom\dagger}={\rm e}^{i2\theta} \sigma^-_j$ for any
$\theta$, thus exhibiting $U(1)$ symmetry and conservation of the
number of excitons $N_{exc} = N_{ph} + 2N_{q}$ where $N_q$ is the
number of qubits in the excited state and $N_{ph}$ is the number pf
photons. However, the CR terms break the $U(1)$ symmetry, they pick up
a phase ${\rm exp}(i4\theta)$ due to the action of the operator ${\cal
  R}(\theta)$; this reduces the symmetry to $Z_4$. Now the system is
left invariant by this rotation only for $\theta=n2\pi/4$, $n=0,1,2,3$
and $N_{exc}$ is no longer conserved.  The discrete $Z_4$ symmetry can
break spontaneously in the ground state leading to an ordered BEC
phase.

To characterize the various possible phases, we calculate several Green 
functions,
\begin{equation}
 G_{\alpha,\beta}(r) \equiv \frac{1}{2N} \sum_i \langle \alpha_i \beta_{i+r} + 
{\rm h.c.} \rangle,
\label{greens}
\end{equation}
where $\alpha$ and $\beta$ denote creation and annihilation operators
of the photons (${\hat a}^\dagger_i$ and ${\hat
  a}^{\phantom\dagger}_i$) or the qubits ($\sigma^+_j$ and
$\sigma^-_j$). $\langle {\cal O}\rangle$ denotes the ground state
expectation value, $\langle GS |{\cal O} | GS \rangle$, for DMRG, and
the statistical average for QMC,
  \begin{equation}
    \langle {\cal O}\rangle \equiv \frac{1}{Z(\beta)}{\rm Tr} \left
      [{\rm e}^{-\beta H}{\cal O}\right ],
  \end{equation}
where $Z={\rm Tr} \left [{\rm e}^{-\beta H}{\cal O}\right ]$ and
$\beta = 1/T$. For example, the one-body photon Green function at
equal time is given by,
\begin{equation}
 G_{a^\dagger, a^{\phantom\dagger}}(r) = \frac{1}{2N} \sum_i \langle 
{\hat a}^\dagger_i {\hat a}^{\phantom\dagger}_{i+r} + {\hat a}^\dagger_{i+r} 
{\hat a}^{\phantom\dagger}_{i} \rangle.
\label{photongreen}
\end{equation}
The qubit Green function is,
\begin{equation}
 G_{\sigma^-, \sigma^+}(r)=\frac{1}{2N}\sum_i \langle \sigma^-_i
 \sigma^+_{i+r} + \sigma^-_{i+r}\sigma^+_i \rangle,
 \label{qubitgreen}
\end{equation}
and the following two functions will be particularly useful:
\begin{equation}
  G_{\sigma^+, a}(r)=\frac{1}{2N}\sum_i \langle {\hat a}^{\phantom\dagger}_i 
\sigma^+_{i+r} 
+ \sigma^-_{i+r}{\hat a}^\dagger_i \rangle,
\label{qubitphoton}
\end{equation}
and
\begin{equation}
  G_{\sigma^+, a^2}(r)=\frac{1}{2N}\sum_i \langle {\hat a}^{2}_i \sigma^+_{i+r} 
  + \sigma^-_{i+r}{\hat a}^{\dagger 2}_i \rangle.
  \label{qubitphoton2}
\end{equation}
The photon pair Green function is given by,
\begin{equation}
  G_{a^{\dagger\,2},a^2}(r)=\frac{1}{N}\sum_i\langle {\hat
    a}^{\dagger\, 2}_{i}a^2_{i+r}+ H.c.\rangle.
  \label{gpair}
\end{equation}
The Green functions described by Eqs. (\ref{greens}-\ref{gpair}) are
defined on a lattice with perdiodic boundary conditions used in the
QMC simulations. When using DMRG with open boundaries, we choose the
origin of the correlation function in the center of the lattice and
calculate the correlations from that point.  Power law decay of one of
these Green functions would indicate quasi-long range order for the
corresponding quantity. If the Green function decays to a finite
constant, it signals long range order and the spontaneous breaking of
the corresponding symmetry. The Fourier transform of the single
particle Green function, Eq.(\ref{photongreen}), gives the photon
momentum distribution, $n_{ph}(k)$; the Fourier transform of the pair
Green function gives the pair momentum distribution, $n_{pair}(k)$;
the Fourier transform of the qubit Green function,
Eq.(\ref{qubitgreen}), gives the qubit momentum distribution,
$n_{q}(k)$ . These quantities will indicate whether a condensate ({\it
  i.e.} long range order) is present. We also measure the average
number of excitons,
\begin{equation}
 N_{exc} = \sum_i \langle {\hat a}^\dagger_i{\hat a}^{\phantom\dagger}_i + 
2\sigma_i^+\sigma_i^- \rangle.
\label{nexc}
\end{equation}
We calculate these quantities using the ALPS DMRG package\cite{alps}
and the SGF QMC algorithm\cite{rousseau08,rousseau08b}. For DMRG, we
verified that, in all cases, the number of states we kept and sweeps
we performed were sufficient for proper convergence (up to $240$
states and $300$ sweeps for large systems).

\section{Results}

\begin{figure}[!ht]
\includegraphics[width=1
  \columnwidth]{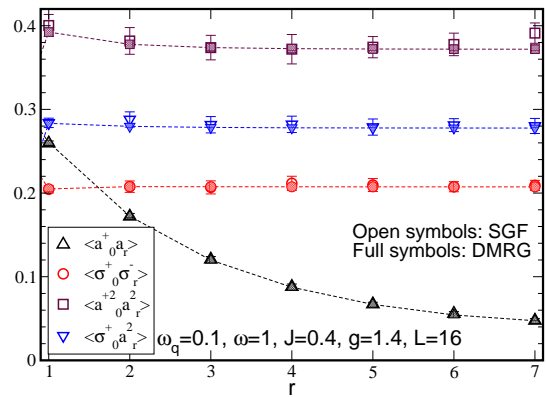}
\caption{(Color online) Green functions versus distance in the pair
  superfluid phase showing excellent agreement between the results of
  the SGF QMC simulations (at $\beta=360$) and the zero-temperature
  DMRG calculations. Here as throughout the paper we take $U=1$.}
\label{dmrgsgfcompare}
\end{figure}

The results we present here were obtained using DMRG and SGF QMC
depending on the physical quantities being studied and the boundary
conditions: Open boundary conditions were typically used with DMRG and
periodic ones with SGF. We begin, therefore, by comparing the two
methods to ensure that results yielded by one can be reliably compared
with the other. Figure \ref{dmrgsgfcompare} compares DMRG and SGF
results, both with periodic boundary conditions, for several Green
functions in the PSF phase (see below) versus distance. The finite
temperature SGF QMC simulations were performed at very low
temperature, $\beta=360$, and show excellent agreement with the
zero-temperature DMRG calculations.

\begin{figure}[!ht]
\includegraphics[width=1
  \columnwidth]{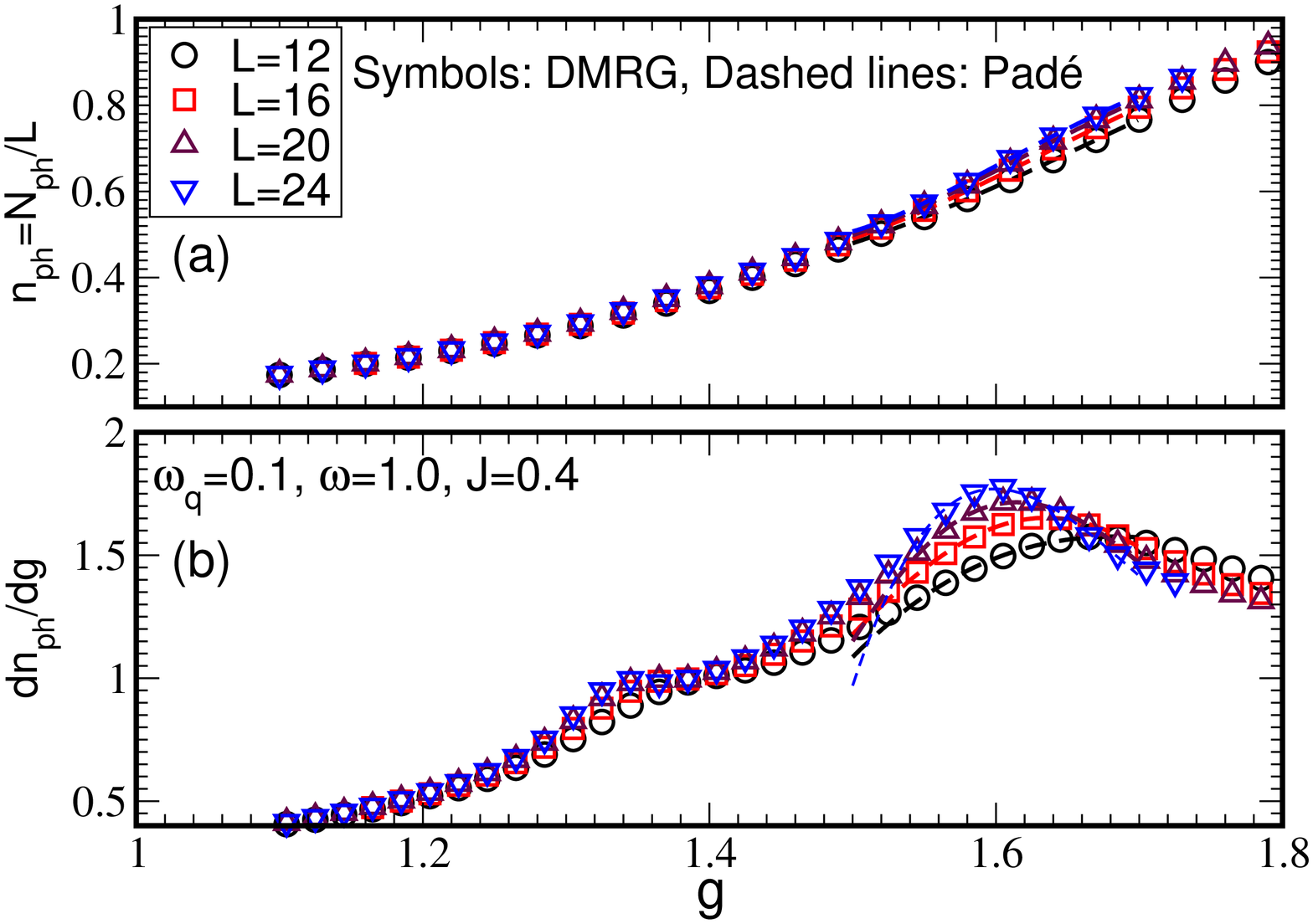}\\
\vskip -0.5cm
\includegraphics[width=1
  \columnwidth]{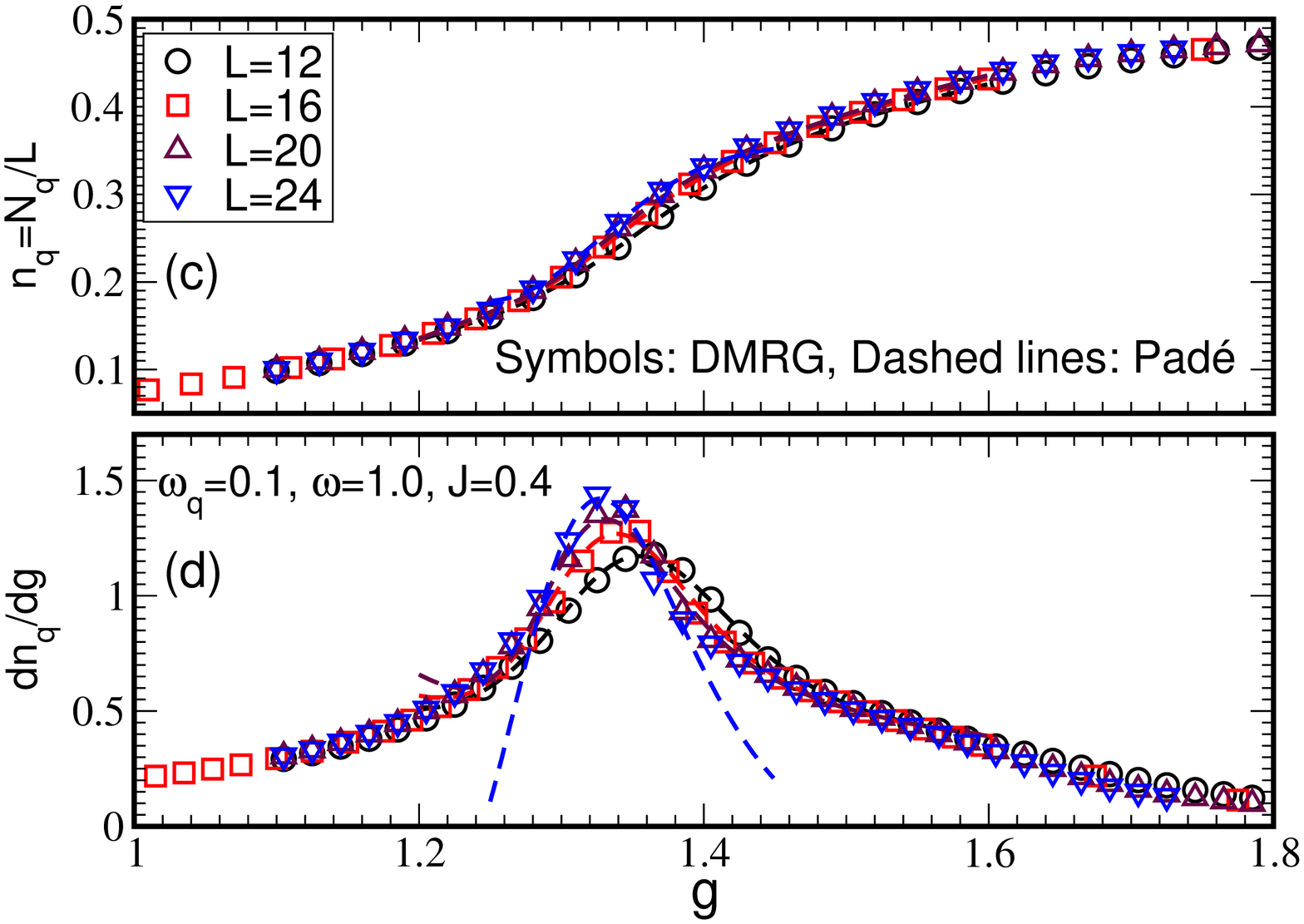}
\caption{(Color online) A cut as a function of $g$ at fixed $J=0.4$
  showing the behavior of (a) the average number of photons per site,
  $n_{ph}$, (b) its derivative ${\rm d}n_{ph}/{\rm d} g$, (c) the
  average number of excited qubits per site, $n_q$, and (d) its
  derivative, ${\rm d}n_{q}/{\rm d} g$. The symbols are results of
  DMRG calculations. The dahsed lines in (a) and (c) are from Pad\'e
  fits to the data, and in (b) and (d) the dashed lines are the
  derivatives of those fits. The symbols in (b) and (d) are numerical
  derivatives of (a) and (c). ${\rm d}n_{ph}/{\rm d} g$ exhibits two
  peaks exposing two quantum phase transitions, ${\rm d}n_{q}/{\rm d}
  g$ exhibits one peak coinciding with the first photon peak. See
  text. Note: The numercial results were obtained with a finer grid
  than shown in the figure. We show only a third of the points to
  improve visibility.}
\label{nphnqvsgJ0.4}
\end{figure}

In what follows, we take $U=1$ to fix the energy scale. We also choose
the photon frequency $\omega=1$ and the qubit energy spacing
$\omega_q=0.1$ and determine the phase diagram of the system governed
by Eq.(\ref{rabihub}) in the $(J,g)$ plane. It was seen in
Refs.\cite{garbe17,cui19} that the average number of excited qubits
per site, $n_q=N_q/L$, is a good probe for the phase transition
between the normal (incoherent) phase and the superradiant
phase. Here, we will use both $n_q$ and $n_{ph}=N_{ph}/L$ (the average
number of photons per site) as probes for possible phase
transitions. We use Pad\'e approximants\cite{cui19} to fit our
numerical results for $n_q$ and $n_{ph}$ as functions of $g$ or
$J$. Differentiating the Pad\'e forms, and also numerically
differentiating the DMRG data, reveal the presence of peaks indicating
the location of the phase transitions. The analytical Pad\'e and the
numerical derivatives are in excellent agreement. Figure
\ref{nphnqvsgJ0.4} shows DMRG results for $n_{ph}$ and $n_q$ and the
corresponding numerical and Pad\'e derivatives, ${\rm d}n_{ph}/{\rm
  d}g$ and ${\rm d}n_{q}/{\rm d}g$, versus $g$ for constant $J=0.4$
and for $L=12,\,16,\,20,\,24$. In Fig.\ref{nphnqvsgJ0.4}(a) we show
$n_{ph}$ versus $g$ at fixed $J=0.4$, and in (b) we show ${\rm
  d}n_{ph}/{\rm d}g$ which, remarkably, displays two peaks indicating
the presence of two quantum phase transitions. $n_q$ and its
derivative are shown in Fig.\ref{nphnqvsgJ0.4}(c) and (d)
respectively. It is seen that ${\rm d}n_{q}/{\rm d}g$ exhibits only
one peak which matches the first peak observed for ${\rm d}n_{ph}/{\rm
  d}g$. We conclude, therefore, that the qubits have one type of
potential order while the photons have two. This will be examined
below. It is seen that the positions of the peaks shift to smaller $g$
values as $L$ increases; this shift is used to extrapolate the
location of the transition to the thermodynamic limit and construct
the phase diagram.

\begin{figure}[!ht]
\includegraphics[width=1
  \columnwidth]{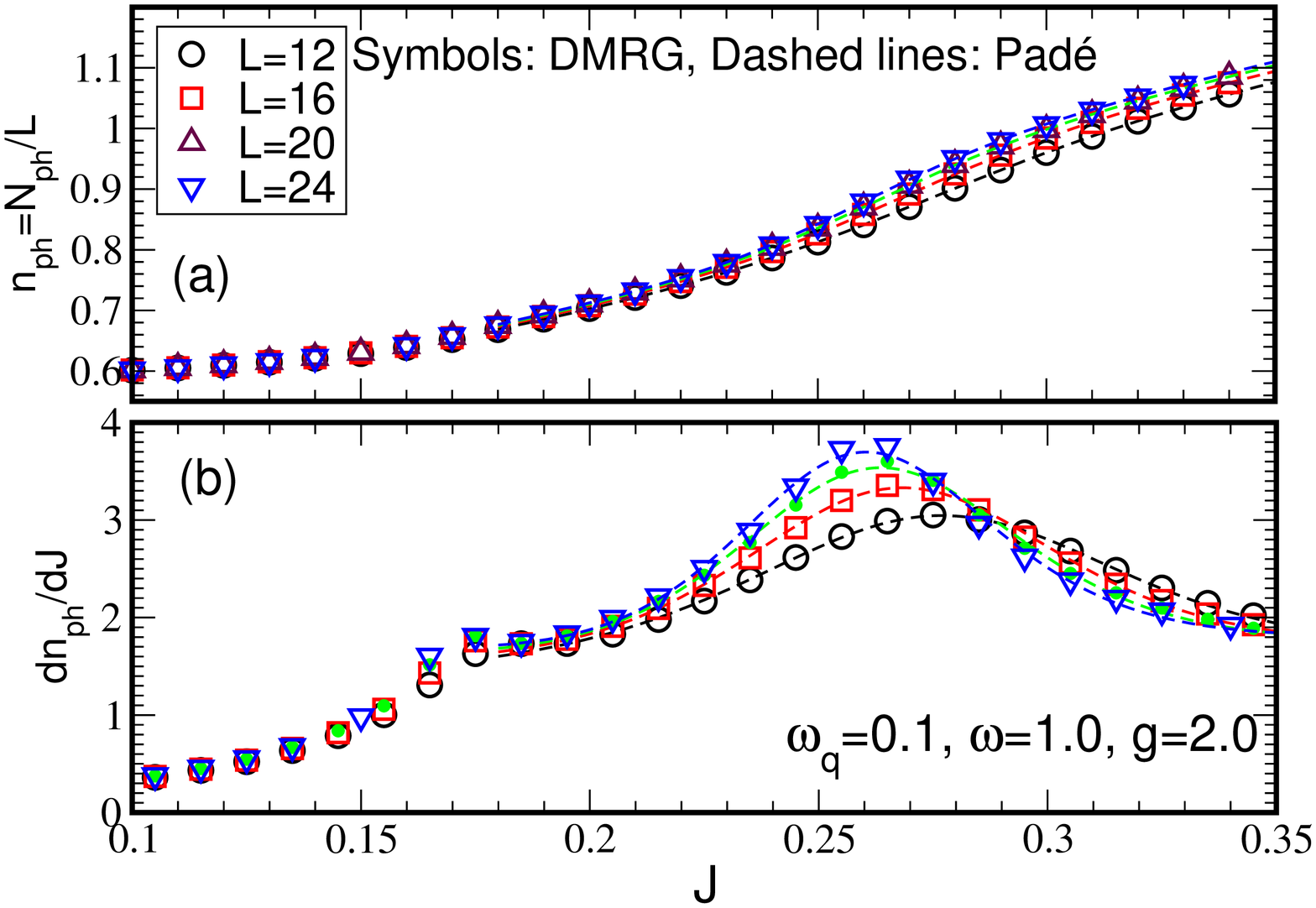}\\
\vskip -0.5cm
\includegraphics[width=1
  \columnwidth]{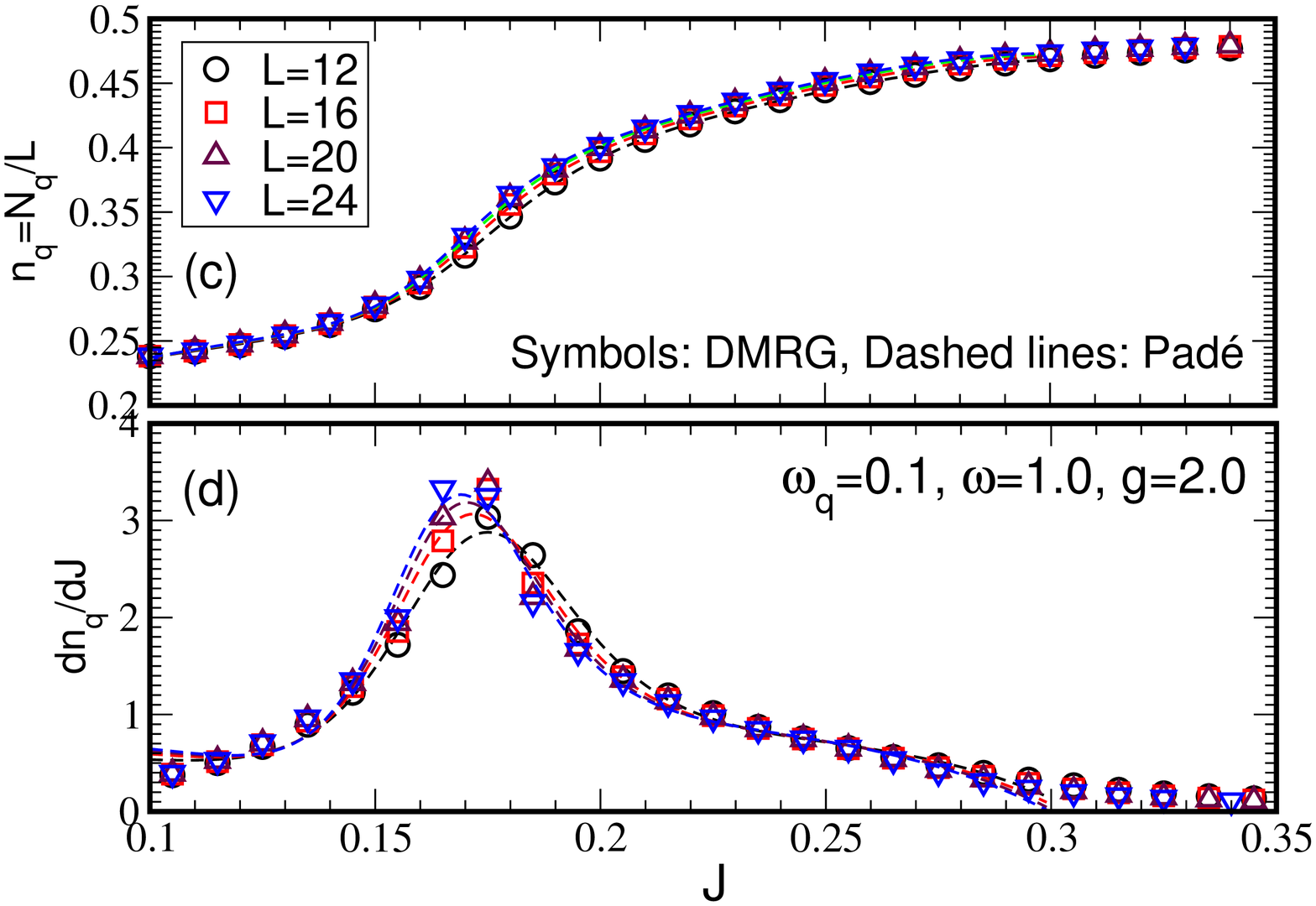}
\caption{(Color online) Similar to Fig.\ref{nphnqvsgJ0.4} but here we
  fix $g=2.0$ and vary $J$.}
\label{nphnqvsJg2.0}
\end{figure}

Similar behavior is observed when $g$ is fixed and and $J$ is changed,
as shown in Figs.\ref{nphnqvsJg2.0}.
We conclude, therefore, that typically, for this system, two quantum
phase transitions are encountered as $g$ ($J$) is fixed and $J$ ($g$)
is varied. We recall that for the two-photon Rabi model in the absence
of the quartic term, no phase transition is present\cite{cui19}. The
addition of the quartic term, $U\sum_i (a^\dagger_i
a^{\phantom\dagger}_i)^2$, stabilized the system and allowed the
appearance of quantum phase transitions.

\begin{figure}[!ht]
\includegraphics[width=1
  \columnwidth]{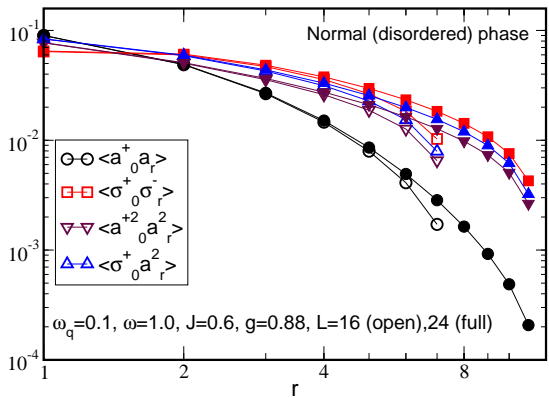}
\caption{(Color online) Four Green functions (DMRG) versus distance
  for $L=16,\,24$. Results for $L=12,\,20$ show the same behavior but
  are not shown to keep the figure uncluttered. The decay with
  distance is faster than power (exponential) indicating a disordered
  phase.}
\label{greensnormalphase}
\end{figure}

To determine the nature of the three phases present, we study the
behavior of the Green functions,
Eqs.(\ref{greens}-\ref{gpair}). Figure \ref{greensnormalphase} shows
four of these functions, obtained with DMRG, on log-log scale clearly
exhibiting decay faster than power (exponential) with distance and,
therefore, corresponding to the normal (incoherent) phase. This
behavior is characteristic of the region in parameter space, ($J,g$),
where $J$ and/or $g$ are small, {\it i.e.} before the first peak in
Figs.\ref{nphnqvsgJ0.4} and \ref{nphnqvsJg2.0} leading to the
conclusion that this region corresponds to a disordered (incoherent)
phase.

\begin{figure}[!ht]
\includegraphics[width=1
  \columnwidth]{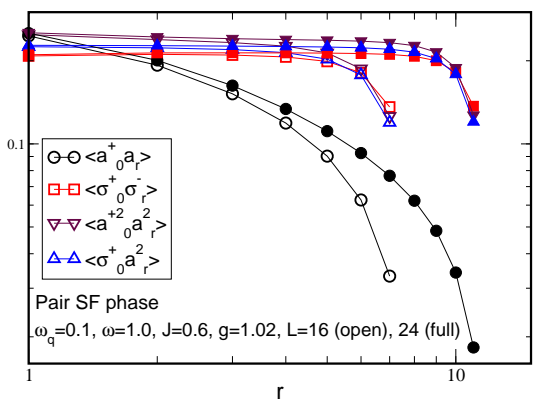}
\caption{(Color online) The photon one-body Green funcion (DMRG)
  decays exponentially while the others saturate to a finite value
  indicating the presence of pair SF but not single particle SF. }
\label{greensPSF}
\end{figure}

Choosing, for example, $(J,g)=(0.6,1.02)$ puts the system between the
two peaks observed in Figs.\ref{nphnqvsgJ0.4} or
\ref{nphnqvsJg2.0}. We see in Fig.\ref{greensPSF} that, in this case,
the one-body photon Green function, Eq.(\ref{photongreen}), still
decays exponentially whereas all other Green functions shown, rapidly
saturate to constant values. The exponential decay of
$G_{a^\dagger,a}(r)$ indicates the absence of off-diagonal long range
or quasi-long range photon order and, consequently, no photon SF, and
$\langle \hat a\rangle=0$. However, the saturation (constant value) of
the other Green functions as the distance increases, indicates the
establishment of true off-diagonal long range order (ODLRO) for {\it
  photon pairs}. The saturation value is the square of the order
parameter, {\it e.g.}  $G_{a^{\dagger\,2},a^2}^{saturated} \sim
|\langle a^2\rangle|^2$. Therefore, between the two peaks of
Figs.\ref{nphnqvsgJ0.4} and \ref{nphnqvsJg2.0}, the system is in a
photon pair SF (PSF) phase with a pair BEC. In terms of symmetry
breaking, this means that the $Z_4$ symmetry of the Hamiltonian,
Eq.(\ref{rabihub}), generated by the operator Eq.(\ref{rotop}), is
only partially broken: $\langle {\hat a}^2\rangle \neq 0$, and
$\langle {\hat a}\rangle =0$.

\begin{figure}[!ht]
\includegraphics[width=1
  \columnwidth]{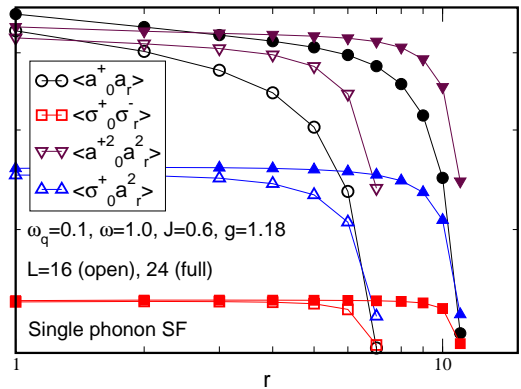}
\caption{(Color online) All the Green functions (DMRG) tend to a
  finite value as the distance increases. This shows that the $Z_4$
  symmetry is completely broken and the system is in the single
  particle SF phase with a photon BEC.}
\label{greensSPSF}
\end{figure}

Increasing $g$ and/or $J$ further, puts the system on the right of the
second peaks in Figs.\ref{nphnqvsgJ0.4} and \ref{nphnqvsJg2.0}.  We
see in Fig.\ref{greensSPSF} that now all Green functions saturate to
constant values as the distance increases. This shows that now the
system is in a phase with true off-diagonal long range photon order
indicating the complete breaking of the $Z_4$ symmetry and the
establishment of a single photon SF.

\begin{figure}[!ht]
\includegraphics[width=1 \columnwidth]{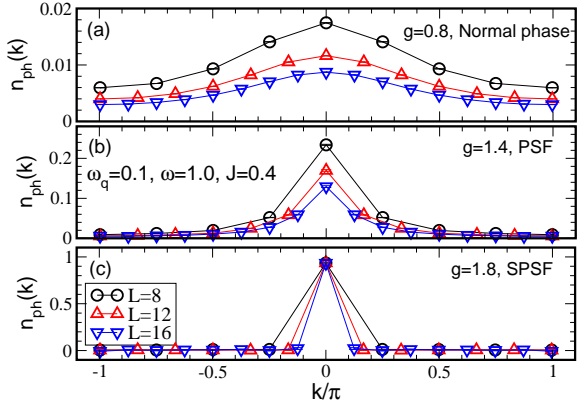}
\caption{(Color online) The normalized single particle momentum
  distribution, $n_{ph}(k)$ for $L=8,\,12,\,16$ in the normal, PSF and
  SPSF phases. The photons form a condensate only in the SPSF phase
  where $n_{ph}(k=0)$ remains constant as $L$ increases. The data were
  obtained with QMC.}
\label{momdistSP}
\end{figure}

To elucidate further the nature of the two SF phases, we examine the
behavior of the qubit, the one-body and the pair momentum
distributions. In Fig.\ref{momdistSP} we show the normalized photon
momentum distribution, $n_{ph}(k)$, in the three phases we have
identified and for several system sizes. We see that, as the system
size increases, $n_{ph}(k=0)$ decreases when the system is in the
normal or PSF phases indicating the absence of a photon BEC and,
therefore, the absence of (ODLRO), $\langle {\hat a}_i\rangle =
0$. However, $n_{ph}(k=0)$ remains constant in the SPSF phase
indicating that here the photons have formed a BEC, $\langle {\hat
  a}_i \rangle \neq 0$, and the $Z_4$ symmetry is broken.

\begin{figure}[!ht]
\includegraphics[width=1 \columnwidth]{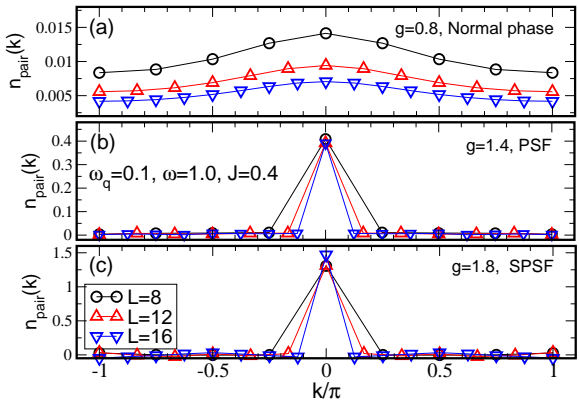}
\caption{(Color online) The normalized pair momentum distribution,
  $n_{pair}(k)$. $n_{pair}(k=0)$ remains constant in the PSF phase as
  $L$ increases signaling the presence of pair BEC. In the SPSF, the
  presence of the BEC is a trivial consequence of the condensation of
  the photons themselves. The data were obtained with QMC.}
\label{momdistPair}
\end{figure}

On the other hand, Fig.\ref{momdistPair} shows the pair momentum
distribution, $n_{pair}(k)$, for the same parameters as in
Fig.\ref{momdistSP}.  Here we see that in the normal phase,
$n_{pair}(k=0)$ decreases as $L$ increases but remains constant in the
PSF phase. This means that in this phase, {\it bound photon pairs}
have formed a BEC even though the photons themselves have not, and,
therefore, $\langle {\hat a}^2_i\rangle \neq0$ but $\langle {\hat
  a}_i\rangle =0$. In the SPSF phase, we already saw,
Fig.\ref{momdistSP}, that $n_{ph}(k=0)\neq 0$, and, therefore, it is
not surprising that $n_{pair}(k=0)$ is also (trivially) nonzero in
this phase.

\begin{figure}[!ht]
\includegraphics[width=1 \columnwidth]{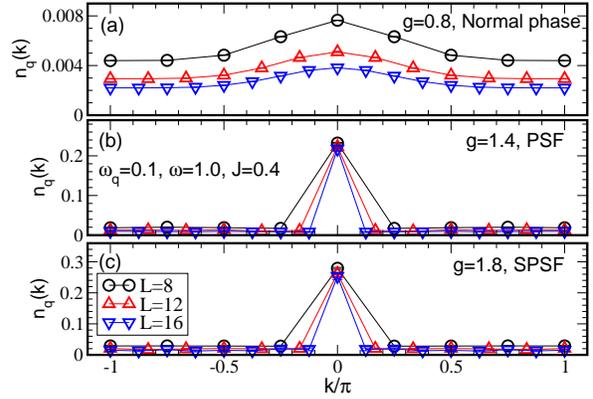}
\caption{(Color online) The normalized qubit momentum
  distribution. The qubits undergo only one transition and,
  consequently, $n_q(k=0)$ behaves the same way in the PSF and SPSF
  phases as evidenced by the closeness of the values. The data were
  obtained with QMC.}
\label{momdistQ}
\end{figure}

Figure~\ref{momdistQ} shows the momentum distribution of the qubits. In
the disordered phase, $n_q(k=0)$ decreases as $L$ increases showing
that there is no order. In both the PSF and SPSF, $n_q(k)$ behaves in
the same way: $n_q(k=0)$ remains constant as $L$ increases showing
that the qubits are ordered. In fact, when the qubits order at the
transition between the normal and PSF phases, they remain ordered in
the same manner as the system transitions from PSF to SPSF. This is
manifested by appearance of only one peak in the derivatives of
$n_{q}$ as seen above.

\begin{figure}[!ht]
\includegraphics[width=1 \columnwidth]{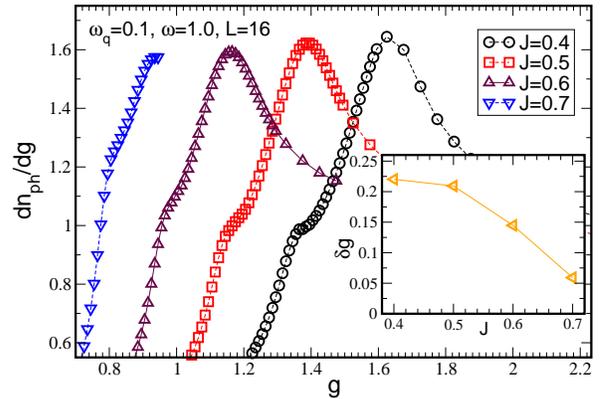}
\caption{(Color online) The main panel shows the two peaks of ${\rm
    d}n_{ph}/{\rm d}g$ approaching each other as $J$ increases. The
  inset shows the separation between the peaks, $\delta g\equiv
  g_c^{(2)}-g_c^{(1)}$, where $g_c^{(1)}$ is the critical $g$ for the
  transition between the disordered and PSF phases, and $g_c^{(2)}$ is
  the critical $g$ for the transition between the PSF and
  SPSF. $\delta g\to 0$ as $J$ increases. Similar behavior is seen for
  ${\rm d}n_{ph}/{\rm d}J$ as $g$ increases. The data were obtained
  with DMRG.}
\label{peaksapproch}
\end{figure}

\begin{figure}[!ht]
\includegraphics[width=1 \columnwidth]{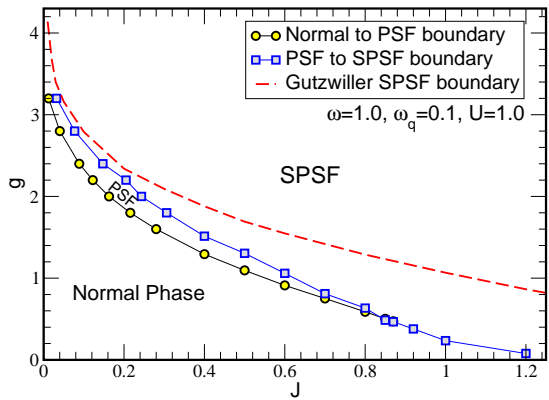}
\caption{(Color online) The phase diagram for $\omega=1$,
  $\omega_q=0.1$ and $U=1$. Three phases are observed, a normal
  disordered phase, the pair superfluid phase (PSF), and the single
  particle superfluid phase (SPSF). We also show (dashed red line) the
  boundary of the SPSF phase obtained with the Gutzwiller mean field
  (see appendix).}
\label{phasediag}
\end{figure}

To map out the phase diagram, we perform simulations along several
lines of constant $g$ while varying $J$ and also constant $J$ while
varying $g$. As shown above, the two peaks exhibited by ${\rm
  d}n_{ph}/{\rm d}g$ and by ${\rm d}n_{ph}/{\rm d}J$ indicate the
location of the quantum phase transition for the system size being
studied. We do the calculations for several system sizes
($L=12,\,16,\,20,\,24$) and extrapolate the critical values to the
thermodynamic limit. The separation between the two peaks in the
derivatives of $n_{ph}$ is not constant; it decreases as $g$ or $J$
get larger. Figure \ref{peaksapproch} illustrates this for several
values of $J$. The inset shows the separation between the two peaks,
$\delta g\equiv g_c^{(2)}-g_c^{(1)}$, where $g_c^{(1)}$ is the
critical $g$ for the transition between the disordered and PSF phases,
and $g_c^{(2)}$ is the critical $g$ for the transition between the PSF
and SPSF. We see that $\delta g$ gets smaller as $J$ increases. We
observe similar behavior for ${\rm d}n_{ph}/{\rm d}J$ as $J$
increases.

Putting all these results together leads to the phase diagram we show
in Fig.\ref{phasediag}. The figure shows the three phases discussed
above: Normal incoherent phase, superradiant photon condensate phase
(single particle SF, SPSF) and sandwiched in between is the pair SF
phase (PSF). As the two transitions approach each other, for example
at large $J$, they become hard to distinguish and appear to merge
eventually into one single transition leading to a direct passage from
the normal phase to the SPSF phase without passing first through the
PSF phase. We also show in Fig.\ref{phasediag} the Gutzwiller mean
field result for the boundary between the SPSF and the other phases
(see appendix for details).

\section{Conclusions}

A lot of attention has been given to the two-photon Rabi model, its
applications \cite{bertet02,stufler06,delvalle10,felicetti18} and the
spectral collapse
\cite{emary02,dolya09,travenec12,maciejewski15,travenec15,peng17,chen12,felicetti15}
it undergoes in certain regions of its parameter space. Interestingly,
in contradistinction with the one-photon Rabi-Hubbard model which, in
its ground state, exhibits a quantum phase transtion from a disordered
phase to a superradiant one, chracterized by a spontaneously broken
symmetry ($Z_2$) and a photon BEC, the two-photon RH model only
exhibits the disordered phase. In this work, we used exact
computational methods (DMRG and QMC) and showed that the system can be
stabilized by a nonlinear (quartic) term which models effective
photon-photon interactions \cite{gorlach17,mittal18,olekhno19}. This
stabilization eliminates the spectral collapse of the model and, in
fact, exposes two quantum phase transtions in the ground state. The
first transition, a consquence of partial spontaneous symmetry
breaking, takes the system from the disordered phase to the pair
superfluid (PSF) phase which is characterized by a nonvanishing pair
condensate order parameter, $\langle a^2\rangle\neq 0$, while at the
same time $\langle a\rangle =0$. The second transition is from the PSF
to the single particle SF (SPSF) phase and completes the symmetry
breaking with the photon condensate order parameterm acquiring a
nonvanishing expectation value, $\langle a\rangle \neq 0$.

An interesting question to ask is what happens in higher order photon
processes, for example the three-photon model? Without stabilization,
the three-photon model undergoes spectral collapse but, with a
nonlinear photon-photon term like we used here, it will be
stabilized. It would be interesting to study this model both in its
Jaynes-Cummings limit (ignoring the counter-rotating terms) where the
symmetry is $U(1)$ and will not break in one dimension, and also in
its full Rabi form where the CR terms are kept and where the symmetry
is now $Z_6$. Specifically, will there be two or more kinds of SF
phases? For example, will there be photon triplet SF and BEC where
three photons act as a single boson that condenses? Will the $Z_6$
symmetry break down in more stages than the smaller $Z_4$ and,
consequently, result in more phases?

%%%%%%%%%%%%%%%%%%%%%%%%%%%%%%%%%%%%%%
\begin{acknowledgments}
S. C. and W. G. are supported by the NSFC under Grant No.~11775021 and
No.~11734002; S. C. is also supported under NSAF Grant No.~U1930402 at
CSRC.
\end{acknowledgments}

\appendix
%%%%%%%%%%%%%%%%%%%%%%%%%%%%%%%%%%%%%%%%%%%%%%%%%%%%%%

\section{Gutzwiller mean field}

We used Gutzwiller mean field to determine the phase boundary of the
single particle superfluid (SPSF), in other words where the order
parameter $\langle \hat{a}_i\rangle=\langle \hat{a}^+_i\rangle=\psi$
acquires a nonzero value, $\psi \neq 0$. Outside the SPSF, $\psi = 0$
both in the pair superfluid phase (PSF) and the disordered phase. We
do not determine the boundary between the disordered phase and PSF
where $\langle a^2_i\rangle $ acquires a nonzero value while $\psi =
0$.

We start by writing
\begin{eqnarray}
{\hat a}^\dagger_i {\hat a}^{\phantom\dagger}_{i+1} \approx\langle
{\hat a}^\dagger_i\rangle {\hat
  a}^{\phantom\dagger}_{i+1}+\langle{\hat
  a}^{\phantom\dagger}_{i+1}\rangle{\hat a}^\dagger_i -\langle {\hat
  a}^\dagger_i\rangle\langle {\hat a}^{\phantom\dagger}_{i+1}\rangle,
\end{eqnarray}
where we ignore fluctuations. This decouples the sites of the lattice
and leads to the single-site Hamiltonian
\begin{eqnarray}
 H_i&=&-2J(\psi{\hat a}^\dagger_i+\psi{\hat
   a}^{\phantom\dagger}_i)+\omega {\hat a}^\dagger_i {\hat
   a}^{\phantom\dagger}_i+\omega_q \sigma^+_i \sigma^-_i \nonumber
 \\ &+&g( \sigma^+_i + \sigma^-_i) ({\hat a}^2_i+{\hat a}^{\dagger
   2}_i)+U({\hat a}^\dagger_i {\hat a}^{\phantom\dagger}_i)^2.
\end{eqnarray}

Now we define the wavefunction basis: $|n_{ph},g\rangle$ and
$|n_{ph},e\rangle$ where $n_{ph}=0,1,2,\dots, n_{max}$ is the number
of photons in the cavity; $g$ ($e$) denotes a qubit in the ground
(exicted) state. The wave function can then be written in terms of
this basis,
\begin{equation}
|\Psi\rangle=\sum_{i=0,\,k=(g,e)}^{n_{max}}c_{k,i}|i,k\rangle.
\end{equation}
The order parameter, $\psi = \langle \Psi | \ a |\Psi \rangle$
becomes,
\begin{eqnarray}
  \nonumber
  \psi&=&c_{g,0}*c_{g,1} + \sqrt{2}c_{g,1}*c_{g,2} \\
  && \ldots+\sqrt{n_{max}}c_{g,n_{max}-1}*c_{g,n_{max}} 
\nonumber \\ &+&c_{e,0}*c_{e,1}+\sqrt{2}c_{e,1}*c_{e,2}+\ldots
\nonumber \\ &+&\sqrt{n_{max}}c_{e,n_{max}-1}*c_{e,n_{max}}
\end{eqnarray}
The coefficients are first chosen randomly, the Hamiltonian matrix is
calculated and diagonalized. This gives a new estimate for the ground
state wavefunction which is then used iteratively to calculate an
improved ground state wavefunction and so on until the process
converges. This way, we calculate $\psi$ for a chosen fixed value of
$U$ and many values of $\omega,\, \omega_q,\, J,\, g$, and determine
the region in phase space where $\psi\neq 0$. For example, for $U=1$,
we obtain the dashed red line in Fig.\ref{phasediag}.

As an example, we show the Hamiltonian matrix for $n_{max}=2$ with
basis order:$|0,g\rangle$, $|1,g\rangle$, $|2,g\rangle$,
$|0,e\rangle$, $|1,e\rangle$, $|2,e\rangle$:
\begin{equation}
\left(\begin{array}{cccccc}
    0 & -2J\psi & 0 & 0 & 0 & \sqrt{2}g \\
    -2J\psi & \omega+U & -2\sqrt{2}J\psi & 0 & 0 & 0 \\
    0 & -2\sqrt{2}J\psi & 2\omega+4U & \sqrt{2}g & 0 & 0 \\
    0  & 0 & \sqrt{2}g & \omega_q & -2J\psi & 0 \\
    0 & 0 & 0 & -2J\psi & \omega +\omega_q+U & -2\sqrt{2}J\psi \\
    \sqrt{2}g & 0 & 0 & 0 & -2\sqrt{2}J\psi & 2\omega+\omega_q+4U
\end{array}\right)
\end{equation}
\hspace{8cm}

%%%%%%%%%%%%%%%%%%%%%%%%%%%%%%%%%%%%%%%%%%%%%%%%%%%%%%%

\end{document}